# Axial magnetic field and toroidally streaming fast ions in the dense plasma focus are natural consequences of conservation laws in the curved axisymmetric geometry of the current sheath

S K H Auluck
Physics Group, Bhabha Atomic Research Center,
Mumbai, 400085 India

## Abstract

Direct measurement of axial magnetic field in the PF-1000 dense plasma focus (DPF), and its reported correlation with neutron emission, call for a fresh look at previous reports of existence of axial magnetic field component in the DPF from other laboratories, and associated data suggesting toroidal directionality of fast ions participating in fusion reactions, with a view to understand the underlying physics. In this context, recent work dealing with application of the hyperbolic conservation law formalism to the DPF is extended in this paper to a curvilinear coordinate system, which reflects the shape of the DPF current sheath. Locally-unidirectional shock propagation in this coordinate system enables construction of a system of 7 one-dimensional hyperbolic conservation law equations with geometric source terms, taking into account all the components of magnetic field and flow velocity. Rankine-Hugoniot jump conditions for this system lead to expressions for the axial magnetic field and three components of fluid velocity having high ion kinetic energy.

# I. Introduction:

One of the reasons why the Dense Plasma Focus (DPF) [1] continues to attract growing interest [2] from plasma physicists is the still-unresolved issue of its much-higher-than-thermal fusion reaction rate [3] and unexpectedly high yields of other nuclear reactions [4,5,6] producing short-lived radioisotopes of possible interest for medical applications. The problem with permanently resolving this issue lies in assembling an overall understanding of some of the published experimental data [7] concerning fast ions participating in these reactions. A significant proportion of such data, dealing with existence of toroidally streaming fast ions and associated axial magnetic field, is available only in hard-to-obtain reports and hence they are mostly ignored in reviews of DPF literature [2,8]. Such widely practiced de-emphasis on these phenomena in literature is understandable, since the original authors themselves chose not to highlight these results prominently or pursue them further. Naturally, most theoretical models of DPF [9-20] also find it expedient to ignore axial magnetic field, azimuthal current and azimuthal mass flow. The present state of understanding of DPF physics is faced, on the one hand by un-refuted published data from multiple reputed laboratories concerning existence of significant phenomena which should find a place in the scientific discourse, and on the other hand by a preponderance of scientific literature which tacitly assumes that these phenomena do not exist. This state of affairs is partly a result of the fact that the original experiments were very tedious and expensive to perform / replicate and partly due to absence of a theoretical framework within which they could be understood.

This stalemate is fundamentally affected by the relatively recent work of Krauz et. al. [21] on measurement of axial magnetic field in the PF-1000 device at Warsaw and its correlation with neutron yield reported by Kubes et. al. [22]: they reinforce credibility of these earlier reports and highlight the need to restore their rightful place in DPF literature. They also suggest that theoretical studies can no longer ignore these phenomena as being outside their scope or being of no consequence in the study of mechanisms for neutron emission. This paper therefore seeks to recapitulate data concerning these phenomena, so that their significance is adequately communicated, and *also to demonstrate that they are a natural consequence of hyperbolic*

*conservation laws in the curved axisymmetric geometry of the DPF plasma current sheath (PCS).*

This demonstration is based on construction of a curvilinear coordinate system, which captures the curved axisymmetric shape of the PCS in an analytical form[23] using the Gratton-Vargas (GV) model [24]. In this reference frame, consisting of coordinate axes along the normal and the tangent to the PCS and along the azimuth, shock wave propagation can be represented as locally unidirectional. Conservation laws for mass, momentum and energy, including contribution from electromagnetic fields, supplemented with magnetic field advection for the ionized state behind the shock wave assumed to obey ideal magnetohydrodynamics (MHD), are formulated under the approximation that physical fields have significant variation only along the normal to the sheath. All three components of magnetic field and fluid velocity are taken into account. This results in a system of 7 one-dimensional hyperbolic conservation law equations with geometric source terms. The mathematical theory of hyperbolic conservation law equations provides a generalized framework for constructing solutions consisting of shock waves, which are moving discontinuities, and rarefaction waves, which describe variation of fields behind the moving discontinuities.

The PCS is associated with a travelling wave solution of this system of 1-D hyperbolic conservation law equations: a moving structure that connects an unperturbed region, containing neutral gas and negligible magnetic field, with a post-shock region containing strongly ionized, magnetized plasma described by a trailing rarefaction wave region (~1-3 cm in thickness according to probe measurements [25]). In this transition zone [22], the electron density, temperature and magnetic field vary over many orders of magnitude over scale-length of a few mm. Corresponding characteristic plasma scale-lengths such as Debye length, various collision mean free paths, collisionless electron and ion skin-depths, Larmor radii of ions and electrons also therefore vary over orders of magnitude; the range of variation of each of these spans values both larger and smaller than the physical scale-length of the transition zone. In this situation, electron momentum convection and Hall Effect cannot be neglected [26] and can give rise to azimuthal current in axisymmetric plasma by the following mechanism [26]: A force on a volume element containing both ions and electrons would produce a larger acceleration in the direction of force for electrons than for ions on account of their smaller mass. But this difference in accelerations cannot lead to unlimited charge separation because of quasi-neutrality. This is possible only if the electron acceleration is *predominantly convective*: electrons and ions would then also move in a direction perpendicular to the direction of force, leading to both a transverse current [26] and transverse mass flow. Momentum flux density from electron momentum convection becomes comparable[26] with the Lorentz force density when the collisionless electron skin depth is comparable with the gradient scale length of physical quantities; this situation occurs in the transition zone between the neutral gas and the strongly ionized plasma. Azimuthal symmetry allows counter-streaming azimuthal flow of electrons and ions without being limited by any opposing forces such as those resulting from charge separation, and without violating overall conservation of charge or of azimuthal momentum. Hall Effect plays a role [26] in this process by providing a mechanism for choosing between $+\hat{\theta}$ and $-\hat{\theta}$ directions with the help of a non-zero but small ambient magnetic field [7]. The transition zone can therefore be expected to contain strong azimuthal sheet currents limited only by conservation laws, generating components of magnetic field other than the azimuthal in the post-shock region, accompanied with azimuthal flow. The curved geometry should also result in a tangential flow as the sheath pushes its way through the neutral gas; this has been reported in the numerical simulations of DPF [10,12]. Both the azimuthal and tangential flows should, in principle, be taken into account in the conservation law equations rather than be neglected at the outset as in most theories of DPF [9-20]. This paper demonstrates that expressions for axial component of post-shock magnetic field, comparable in magnitude with the driving azimuthal magnetic field, and for three components of post-shock flow velocity with kinetic energy density comparable with magnetic energy density, naturally emerge from this exercise.

The next section reviews "forgotten" reports concerning existence of axial magnetic field and toroidal directionality of fast ions responsible for reactions in the DPF. Section III describes construction of the curvilinear coordinate system with the help of the Gratton-Vargas model, with details given in the Appendix. Section IV describes formulation of one-dimensional hyperbolic conservation laws and related equations in the local curvilinear coordinate system. Section V constructs the Rankine-Hugoniot jump conditions and derives some results, including expressions for the three components of fluid velocity, axial magnetic field, azimuthal and tangential electric field as well as relations connecting pressure and specific volume for the post-shock state. Section VI presents summary and conclusions.

## II. Review of early work related to axial magnetic field and directionality of reaction producing ions:

The purpose of this section is to review early work suggesting existence of axial magnetic field in the DPF and revealing important features related to directionality of reaction-producing ions, *with particular emphasis on results not available in easily accessible publications*, in order to show that, despite the decision of their authors not to prominently highlight or pursue them in the then-prevailing circumstances, they represent important scientific contributions deserving their rightful place in scientific literature and are of great relevance to the current resurgence of interest[2] in the dense plasma focus.

The first mention of possible presence of axial magnetic field in the DPF came from an experiment reported in 1979 by Gribkov et. al.[27] in which a laser pulse ablated the center of the copper anode of the Filippov type DPF at the Lebedev Physical Institute in Moscow at an instant much before the pinch stage. Time integrated soft x-ray pinhole photography and 5 frame pulsed ruby laser interferometry revealed that the resulting copper plasma plume emerged with negligible radial expansion. In a subsequent experiment [28], the plume was generated in an off-axis location and was seen to be pushed towards the axis and then to expand axially with negligible radial expansion (see Fig.1). Since the DPF plasma sheath and its associated azimuthal magnetic field had not yet reached the vicinity of the copper plume, the authors suggested that this behavior could be caused only by an axial magnetic field.

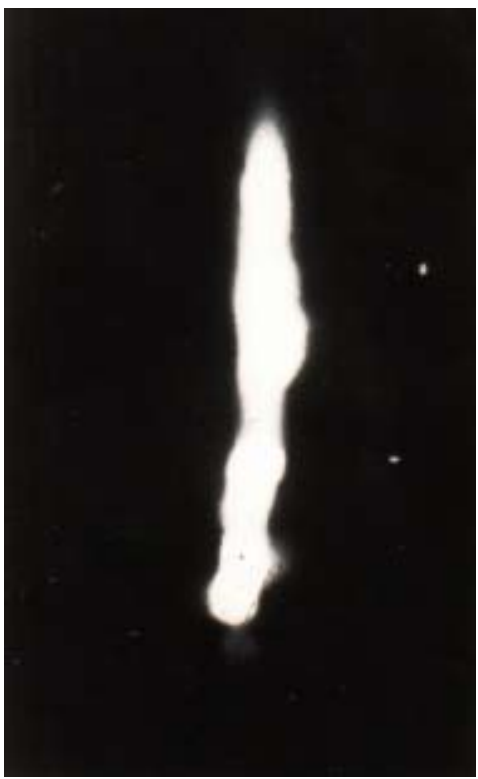

Fig. 1: Time integrated X-ray pinhole picture of a copper plasma plume generated by laser ablation of copper anode of the Fillippov type DPF at the Lebedev Physical Institute. The ablation was at an off-axis location. Although the photograph is time-integrated, rapid radiative cooling of the copper plasma would limit the duration of its x-ray emission to at most a small multiple of the laser pulse duration (5 ns), providing effective time resolution less than the propagation time of the DPF plasma to the center. The x-ray image of the pinch is much less intense than the copper radiation and gets

suppressed in the filtered pinhole image. (Image courtesy Dr. V. A. Gribkov, reproduced with his kind permission)

Axial magnetic field was reported [29] in the Frascati 1 MJ focus using Faraday rotation. Fig 2 shows a schematic of the experiment and the reported result. This report mentions that the observed intensity modulation effect due to Faraday rotation was much larger than the 2° threshold of the optical set-up, and that shadow effect, minimized through depth-of-field control, could not have caused the observed opposite modulation of the two complimentary images of Faraday effect. The report categorically mentions that the polarity of the axial magnetic field was observed to remain constant.

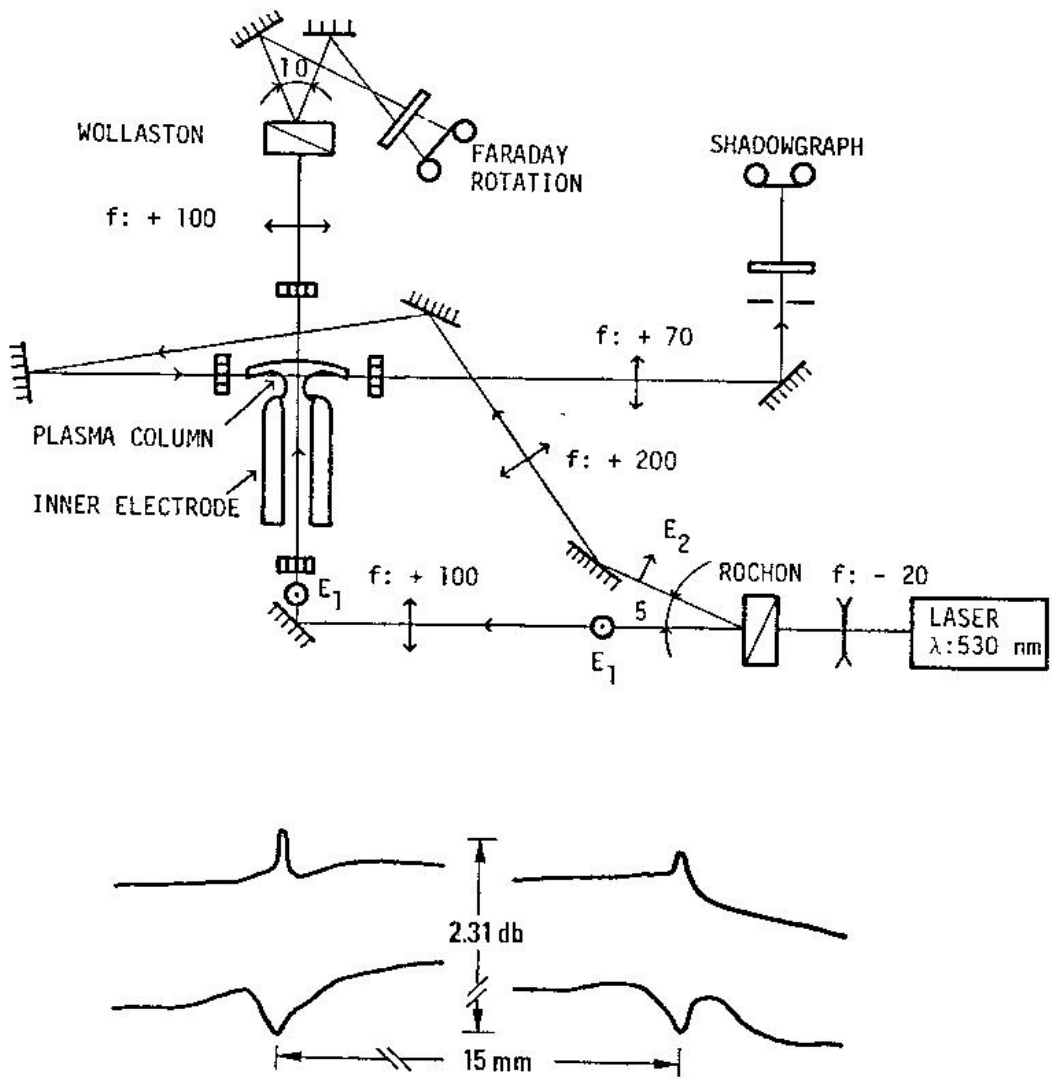

Fig. 2: Detection of axial component of magnetic field by Faraday Effect in the Frascati 1 MJ DPF. (Reproduced from Frascati Report 81.43 with kind permission of Director, ENEA Frascati). The upper part of the figure is the optical schematic; the lower is the densitometric trace on the two complementary images of Faraday effect.

The existence of axial magnetic field in POSEIDON plasma focus was inferred by Jaeger and Herold [30] from distortion and tilting of the reconstructed reaction proton image with respect to the device axis, which could be corrected only by incorporating an azimuthal component of current (~200-600 kA) on the surface of the pinch in their reconstruction algorithm. They also mention that the polarity of this azimuthal current remained constant in about 50 evaluated shots.

Magnetic probes with transverse component rejection ratios better than 30:1 detected [31] signals in the radial implosion phase of the POSEIDON plasma focus at Stuttgart corresponding to radial, azimuthal and axial components of magnetic field having distinct features. A diamagnetic loop placed outside the cathode of a small plasma focus consistently detected a signal [32] indicating rate of change of axial flux being emitted from the plasma.

The realization that the neutron emission from DPF was not of thermal origin came from measurements of neutron *fluence anisotropy* [1]. However, what sort of fast ion population would give rise to observed properties of the neutron emission remains a subject of active research. A recent summary of the status [7] mentions the pioneering observations [33-36] of *axial energy anisotropy* suggesting presence of a linear ion beam along the axis, along with contrary evidence of accelerated ions in the opposite direction [37-39].

Bernard et.al. at Limeil [40] reported in 1975 that the side-on neutron time-of-flight spectrum had much higher width than the end-on spectrum indicating radially streaming deuterons. The neutron time-of-flight measurements in 1978 by Milanese and Pouzo [41] on the 1 MJ Frascati machine using a 128 m long horizontal flight path revealed, for the first time, features which were inconsistent with a beam model of either axial or radial kind: two lateral sub-peaks on either side of a central peak in the neutron energy spectrum could only be explained in terms of loops of 100-keV deuterons in the plane determined by the device axis and the direction of observation.

The possibility that the axial magnetic field may be associated with the mechanism of neutron production was first suggested in 1981 by the Frascati team [29]. They reported experiments indicating presence of azimuthally streaming 100 keV deuterons. Their schematic is shown in Fig 3(a). It consisted of 4 neutron collimators viewing a portion of the pinch zone from left and right sides of the axis from opposite sides. Single-shot neutron spectra recorded on nuclear emulsion plates showed (see Fig 3(b)) that "Irregular patterns of lines have main features centered around 2.2, 2.5 and 2.8 MeV obeying a scheme of deuteron motion circling around the experiment axis".

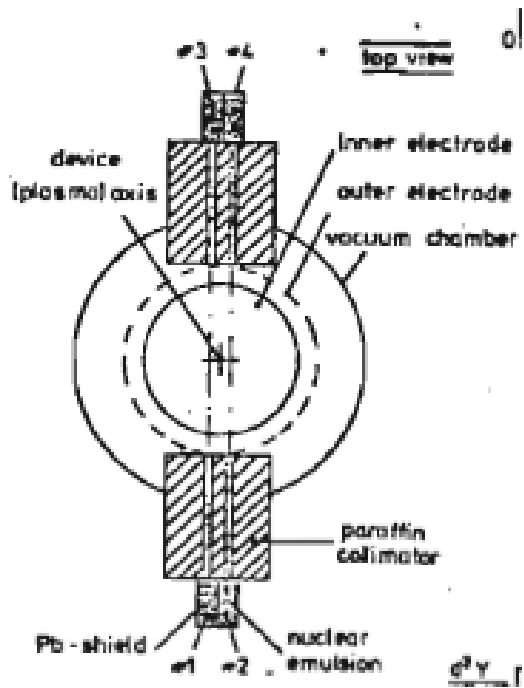
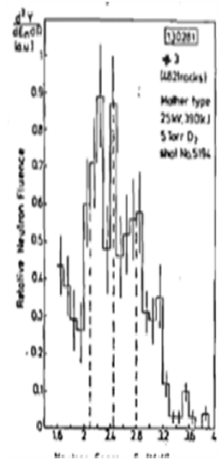
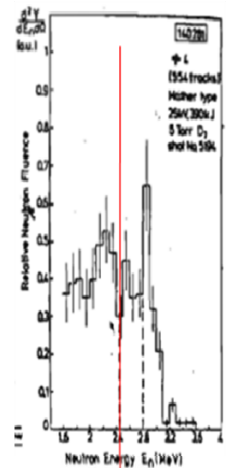
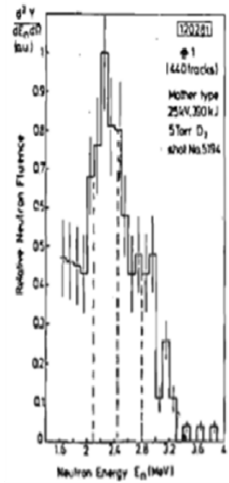
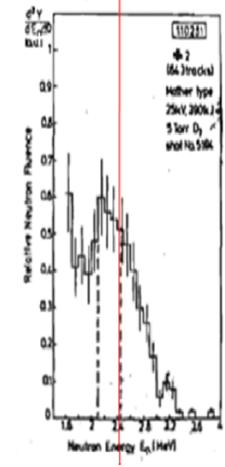

Fig. 3(a) and (b) Reproduced from Frascati Report 81.43 with kind permission from Director, ENEA, Frascati.

The reaction proton spectroscopy work of Jaeger [42] at Stuttgart provided corroboration of the Frascati observations of azimuthally circulating high energy deuterons. Fig. 4(b) shows space-resolved reaction proton spectra corresponding to points marked on the x-ray pinhole photograph in Fig 4(a). The evident asymmetry of spectra from pairs of laterally mirror-symmetric points (#3 and #4, #5 and #6) is a clear indication of presence of azimuthal motion of deuterons participating in the fusion reactions. That such asymmetry is not seen in other pairs of laterally mirror-symmetric points (#1 and #2, #7 and #8) is itself suggestive of a finite axial extent of such azimuthal motion.

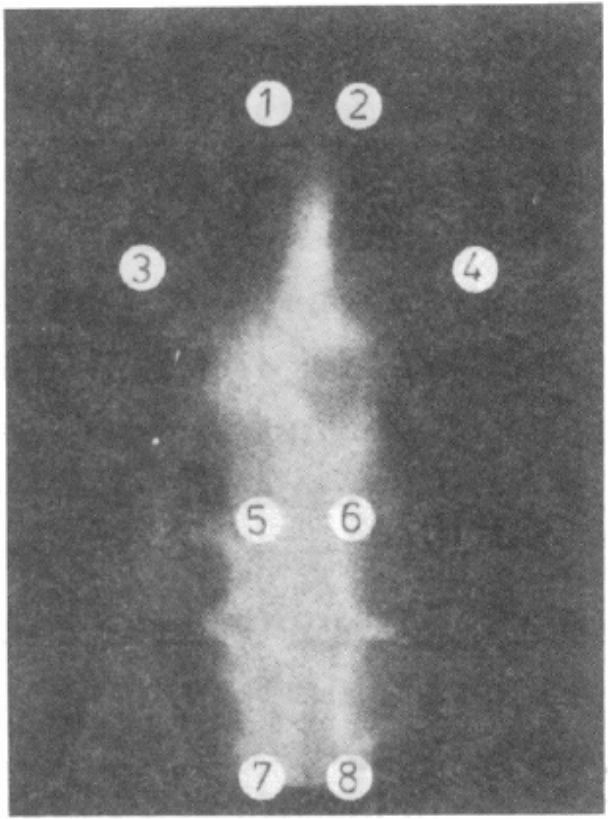
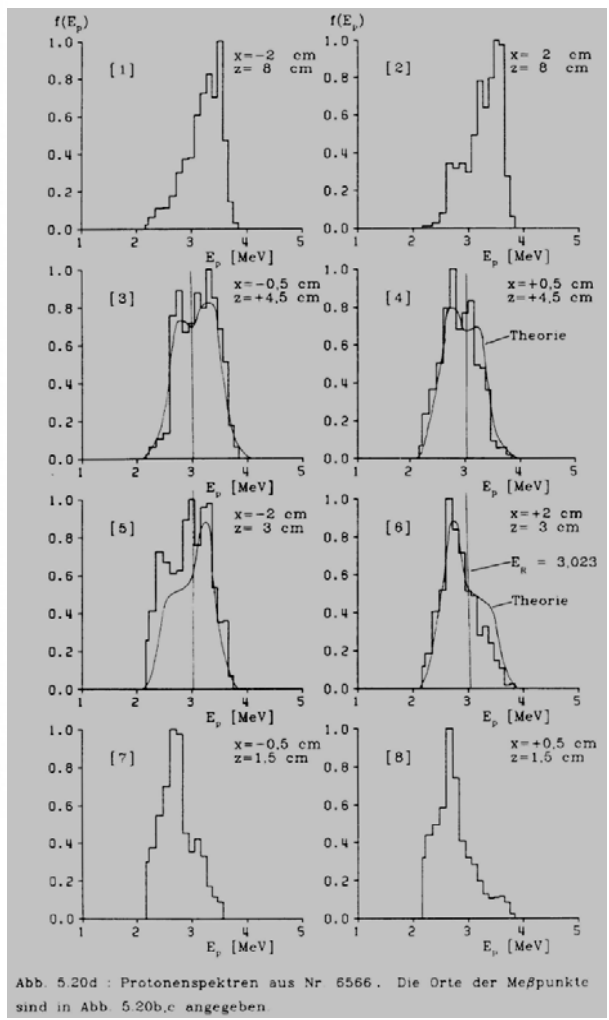

Fig. 4(a) Soft X-ray image showing location of spectra shown in Fig 4(b). Reproduced from the Ph.D. dissertation of U. Jaeger, with kind permission from Director, Institut fuer Plasmaforschung, University of Stuttgart.

The combination of reports of ions moving along the axis[33-36], moving radially [40], moving in loops in the plane comprising device axis and a horizontal line [41], opposite to the axis[37-39] and around the axis [13,42] (over axially limited region) can be comprehensively described as representing toroidal motion [43]. The signature of such toroidal motion is predominantly seen in the width (and not the average energy which remains ~2.45 MeV for neutrons) of the space-averaged side-on neutron spectrum, and in the space space-resolved side-on neutron or fusion proton spectra. Kubes et. al. [44] present interferometric evidence concerning evolution of toroidal and helical structures in PF-1000.

The point of reproducing this data in such detail is to highlight the care, effort and expense invested by leading laboratories of the 1970s and 80s in obtaining these results and to suggest that they represent real physical effects (and not artifacts), which need to be included in the scientific discourse on the dense plasma focus, such as literature reviews, and development of theoretical models.

This paper presents an attempt at an inclusive theoretical exercise, which uses the Gratton-Vargas (GV) model [23,24] to construct a local curvilinear coordinate system in which the shock wave can be described as locally unidirectional. It extends recent work [45] on application of hyperbolic conservation law formalism to the DPF.

### III. The local coordinate system in the GV model:

The details of the GV model presented earlier [23] are not repeated here to conserve space; however, some features essential for this discussion, as well as some changes in nomenclature, are recapitulated for the sake of completeness.

The GV model is based on the idea that the plasma current sheath acquires its velocity and shape from a balance between the magnetic pressure driving it into the neutral gas and the "wind pressure" resisting it – this may be looked upon as a "global momentum conservation" boundary condition. This is written in terms of the velocity $v_n$ normal to the sheath, mass density $\rho_0$ of the fill gas, azimuthal component of magnetic field $B_\theta$ :

$$\rho_0 v_n^2 = B_\theta^2 / 2\mu_0 \Rightarrow v_n = B_\theta / \sqrt{2\mu_0 \rho_0} \qquad 1$$

Assuming anode radius a, temporal current profile I(t), introducing dimensionless cylindrical coordinates $\tilde{r} \equiv r/a$ and $\tilde{z} \equiv z/a$ and independent variable $\tau$ given by

$$\tau(t) = \frac{\mu_0}{\pi a^2 \sqrt{2\mu_0 \rho_0}} \int_0^t I(t') dt' \qquad 2$$

the scaled partial differential equation for the surface $\psi(r,z,t) \equiv z - \bar{f}(r,t) = 0$, representing the shape and position of the "effective magnetic piston" embedded within the plasma current sheath, becomes [23]

$$\partial_\tau \psi + \sqrt{(\partial_{\tilde{r}} \psi)^2 + (\partial_{\tilde{z}} \psi)^2} \frac{1}{2\tilde{r}} = 0 \qquad 3$$

The family of characteristic curves, perpendicular everywhere to the integral surface of 3, (a surface of revolution in 3-D, called the GV surface[23]), is described by the equation

$$\frac{\tilde{z}}{N} + s\mathrm{ArcCosh}\left(\frac{\tilde{r}}{|N|}\right) = C_1 = \mathrm{Constant} \qquad 4$$

Here, N is an invariant of the characteristic curve related to the initial condition of the problem $N = \tilde{r}_i \cos\phi_i \equiv N_i$, where $\phi_i$ is the angle made by the normal to the initial GV surface $\tilde{z}_i = \overline{f}_i(\tilde{r}_i, \tau_i)$ with the z-axis and $\tilde{r}_i$ is the normalized radial coordinate of the intersection of the characteristic curve with the initial GV surface. The variable $s = \pm 1$ represents the sign of $\left(d\overline{f}_i(\tilde{r}_i, \tau_i)/d\tilde{r}_i\right)$.

The GV surface is represented [23] in $(\tilde{r}, \tilde{z})$ space by the parametric curve $\tilde{r} = N\mathrm{Cosh}(\alpha/2)$; $\tilde{z} \equiv N(C_1 - s\alpha/2)$ connecting the anode (or insulator) at normalized anode radius 1 (or normalized insulator radius $\tilde{r}_I$) to the cathode at normalized cathode radius $\tilde{r}_C$, with the value of $\alpha(\tau)$ for any $\tau$ found from the following equation:

$$\mathrm{Sinh}(\alpha(\tau)) + \alpha(\tau) = 2\left(C_2(\tilde{r}_i, \tau_i, N_i) - \frac{s\tau}{N_i^2}\right) \qquad 5$$

$$C_2 = \frac{\tilde{r}}{N}\sqrt{\frac{\tilde{r}^2}{N^2} - 1} + \mathrm{ArcCosh}\left(\frac{\tilde{r}}{|N|}\right) + \frac{s\tau}{N^2} = \mathrm{constant} \qquad 6$$

The initial plasma profile $\tilde{z}_i = \overline{f}_i(\tilde{r}_i, \tau_i)$ at any "initial instant" $\tau_i$ provides the values of constants C$_1$ and C$_2$; the boundary condition[23] that the GV surface must meet the anode at right angles ensures that N=1 during the rundown phase of a Mather type DPF and that N varies continuously from 1 to 0 at the vertex at the end of the rundown phase (see fig 5.). Equations 5 and 6 then show how the GV surface propagates as a function of $\tau$ (see fig 1. of Ref 23). The

GV surface resembles the plasma shape seen in the MHD calculations [10,11] as well as in experiments [46,47].

A local coordinate system $(\zeta,\theta,\xi)$ can be constructed (see fig. 5) with unit vector $\hat{\zeta} = s\tilde{r}^{-1}\sqrt{\tilde{r}^2 - N^2}\hat{z} + \tilde{r}^{-1}N\hat{r}$ along the tangent, a unit vector $\hat{\theta}$ along the azimuth and a unit vector along the normal to GV surface defined as $\hat{\xi} \equiv \hat{\zeta} \times \hat{\theta} = -s\tilde{r}^{-1}\sqrt{\tilde{r}^2 - N^2}\hat{r} + \tilde{r}^{-1}N\hat{z}$, pointing in the direction of the sheath velocity. The coordinate differentials in the local frame of reference are related (see Appendix) to the cylindrical coordinate system by the relations

$$d\xi = -\frac{s\tilde{r}d\tilde{r}}{\sqrt{\tilde{r}^2 - N^2}}; \quad d\zeta = \frac{\tilde{r}}{N}d\tilde{r} \qquad 7$$

Integration of 7 gives

$$\xi = \xi_0 - s\sqrt{\tilde{r}^2 - N^2}; \quad \zeta = \zeta_0 + \tilde{r}^2/2N. \qquad 8$$

The integration constants $\xi_0(\tau,N)$ and $\zeta_0(\tau,N)$ are chosen so that the origin of the coordinate system lies on the intersection of the GV surface specified by $\tau$ with the characteristic specified by N. This coordinate system is illustrated in fig 5. Relations 8 can be inverted:

$$N(\xi,\zeta) = (\zeta - \zeta_0) + S\sqrt{(\zeta - \zeta_0)^2 - (\xi - \xi_0)^2}; \quad S = -s\text{Sign}[\xi - \xi_0] \qquad 9$$

$$\tilde{r}^2(\xi,\zeta) = 2(\zeta - \zeta_0)^2 + 2(\zeta - \zeta_0)S\sqrt{(\zeta - \zeta_0)^2 - (\xi - \xi_0)^2} \qquad 10$$

For the region shown to the left of the N=1 characteristic in fig. 5, s =+1, $\xi_0 > 0, \zeta_0 < 0$. Some properties of this coordinate system are given in the Appendix.

The Gratton-Vargas model is able to predict the shape of the plasma as a function of time [46], the variation of rundown time with pressure [46] and with recent inclusion of circuit resistance [23], is also able to reproduce the experimental current waveforms from four large facilities [47] using gas pressure, static inductance, and circuit resistance as fitting parameters.

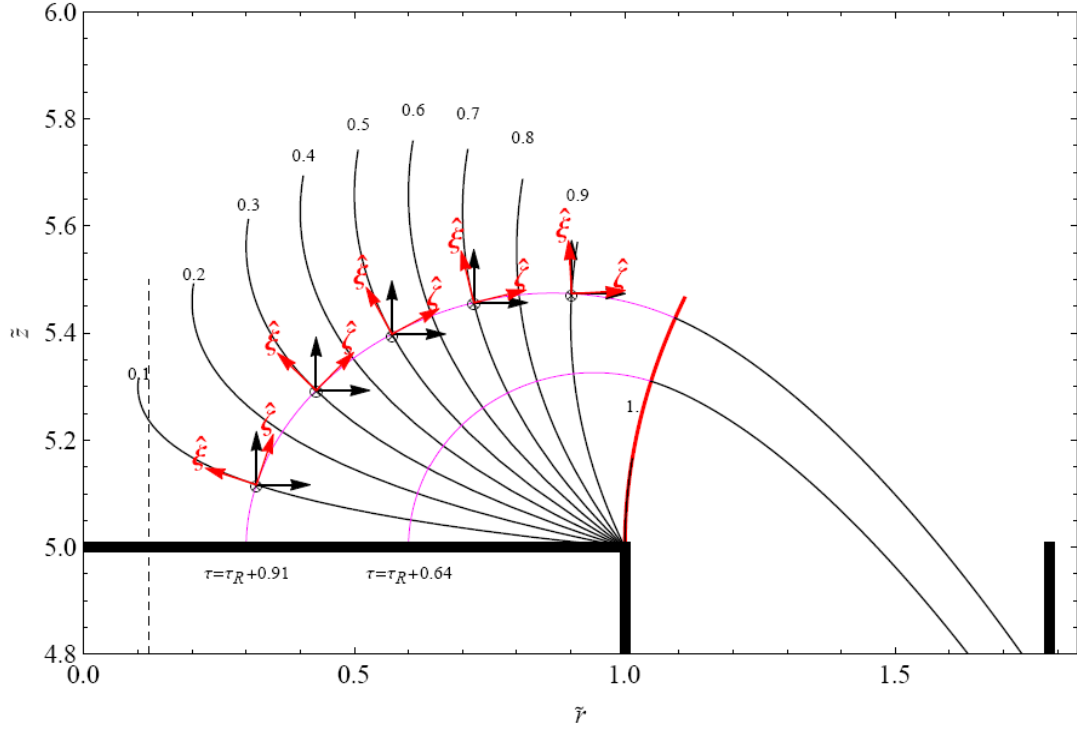

Fig. 5: The GV surface in the turn-around phase is illustrated for two values of $\tau$. A set of 10 characteristic lines drawn from the vertex at the end of anode, labeled by N=0.1 to 1, intersects both GV surfaces. Two local frames of reference, one in cylindrical coordinates $(\tilde{r},\theta,\tilde{z})$ and the other in $(\zeta,\theta,\xi)$ coordinates are drawn with origins at the points of intersection. The configuration of this figure corresponds to that of fig. 1 of Ref 23. However, the definition of unit vectors is different from that in Ref. 23 and integration constants $\xi_0(\tau,N)$ and $\zeta_0(\tau,N)$ are chosen to make the origin coincide with the point of intersection between the characteristic labeled by N and the integral surface labeled by $\tau$. Some vector relations for this system are given in the Appendix.

### IV. Formulation of conservation laws in local coordinate system:

This section begins by recapitulating earlier discussion [45] concerning the conservation of mass, momentum and energy including contributions from the electromagnetic field, described by the following equations 11, in which $\rho$ is the mass density, $\vec{v}$ is the fluid velocity, p is the fluid pressure, $\vec{B},\vec{E}$ are the magnetic and electric fields, $\mathcal{E}$ is the total energy density, $\mathcal{W}_L$ is the loss of energy per unit volume per unit time through processes such as radiation and thermal conduction and $\overset{\leftrightarrow}{\mathbb{I}}$ is a unit dyadic:

$$\partial_t \rho + \vec{\nabla} \cdot (\rho \vec{v}) = 0$$
$$\partial_t \left( \rho \vec{v} + \vec{E} \times \vec{B} / \mu_0 c^2 \right) + \vec{\nabla} \cdot \left( \rho \vec{v} \vec{v} + \mathbb{I} \left( p + B^2 / 2\mu_0 \right) - \vec{B}\vec{B} / \mu_0 \right) = 0 \qquad 11$$
$$\partial_t \mathcal{E} + \vec{\nabla} \cdot \left( \mathcal{E} \vec{v} + p\vec{v} + \vec{E} \times \vec{B} / \mu_0 \right) = -\mathcal{U}_L'$$

This is supplemented by the equation of magnetic flux advection in the strongly ionized post-shock plasma, assumed to be adequately represented by ideal magnetohydrodynamics (MHD) [48], and divergence relation for magnetic field:

$$\partial_t \vec{B} + \vec{\nabla} \cdot \left( \vec{v}\vec{B} - \vec{B}\vec{v} \right) = 0 \qquad 12$$

$$\vec{\nabla} \cdot \vec{B} = 0 \qquad 13$$

The pressure is related to number densities $n_i, n_e, n_n$ of ions, electrons and neutrals, and temperature T in energy units, as

$$p = T \left( n_i + n_e + n_n \right) \qquad 14$$

Electrons, ions (of mass $m_i$) and neutrals (of mass $2m_i$) are assumed to have the same temperature T. Using the degrees of dissociation and ionization, $\delta$ and $\alpha$, the total energy $\mathcal{E}$ per unit volume is given [45] by

$$\mathcal{E} = \left( \alpha \rho \varepsilon_i + \delta \rho \varepsilon_d + p/(\gamma - 1) \right) + \rho v^2 / 2 + B^2 / 2\mu_0 \qquad 15$$

In 15 $\varepsilon_d$ represents [45] the energy per unit mass required to fully dissociate the (usually diatomic) neutral gas and $\varepsilon_i$ represents the energy per unit mass required to convert it into fully ionized mono-atomic plasma. For the strongly ionized, magnetized plasma in the post-shock state, the electric field is given by $\vec{E} = -\vec{v} \times \vec{B}$. The Poynting vector is then given by $\vec{E} \times \vec{B} / \mu_0 = \vec{v} B^2 / \mu_0 - \vec{B}(\vec{v} \cdot \vec{B}) / \mu_0$.

Forming dimensionless operators and quantities $\tilde{\nabla} = a\vec{\nabla}$, $\tilde{B} = \vec{B}/B_0$, $\tilde{\rho} = \rho/\rho_0$, $\tilde{v} = \vec{v}/v_A$, $\tilde{p} = p/p_M$, $B_0 = \mu_0 I(\tau)/2\pi a$, $v_A = B_0 / \sqrt{2\mu_0 \rho_0}$, $p_M \equiv B_0^2 / 2\mu_0$;

$$\tilde{\varepsilon}_i = \varepsilon_i/v_A^2\,;\tilde{\varepsilon}_d = \varepsilon_d/v_A^2\,,\quad \tilde{\varepsilon}_{eff} \equiv \left(\alpha_1\tilde{\varepsilon}_i + \tilde{\varepsilon}_d\right)\,,\quad \tilde{\mathcal{E}} \equiv \mathcal{E}/p_M = \alpha\tilde{\rho}\tilde{\varepsilon}_i + \delta\tilde{\rho}\tilde{\varepsilon}_d + \frac{\tilde{p}}{(\gamma-1)} + \frac{1}{2}\tilde{\rho}\tilde{v}^2 + \tilde{B}^2,$$

$\tilde{\mathcal{U}}_L = \mathcal{U}_L/p_M$ and using 2 as $\partial_t = \partial_t\tau\partial_\tau = a^{-1}v_A\partial_\tau$, equations 11 and 12 can be written in dimensionless form:

$$\partial_\tau\tilde{\rho} + \tilde{\nabla}\cdot(\tilde{\rho}\tilde{v}) = 0 \qquad\qquad 16$$

$$\partial_\tau(\tilde{\rho}\tilde{v}) + \tilde{\nabla}\cdot\left(\tilde{\rho}\tilde{v}\tilde{v} + \mathbb{I}(\tilde{p}+\tilde{B}^2) - 2\tilde{B}\tilde{B}\right) = -\tilde{\rho}\tilde{v}\partial_\tau\left(\log(\tilde{I}(\tau))\right) \qquad 17$$

$$\partial_\tau\tilde{\mathcal{E}} + \tilde{\nabla}\cdot\left(\tilde{\mathcal{E}}\tilde{v} + \tilde{p}\tilde{v} + 2\tilde{v}\tilde{B}^2 - 2\tilde{B}(\tilde{v}\cdot\tilde{B})\right) = -\tilde{\mathcal{U}}_L - 2\tilde{\mathcal{E}}\partial_\tau\log\tilde{I}(\tau) \qquad 18$$

$$\partial_\tau\tilde{B} + \tilde{\nabla}\cdot\left(\tilde{v}\tilde{B} - \tilde{B}\tilde{v}\right) = -\tilde{B}\partial_\tau\log\tilde{I}(\tau) \qquad\qquad 19$$

The Poynting vector is of the order of $v_A^2/c^2$ in the second of equations 11 and can be neglected for shock and material velocities much less than velocity of light. As already mentioned in the Introduction, characteristic plasma scale lengths such as Debye length, various collision mean free paths, collisionless electron and ion skin-depths, Larmor radii of ions and electrons vary over orders of magnitude, spanning values both larger and smaller than the physical scale length of the transition zone. A variety of dissipative phenomena can therefore give rise to a shock discontinuity of thickness much smaller than the characteristic dimensions of the GV surface facilitating a locally-planar approximation for shock propagation [45].

Expressions for some differential operators in the local curvilinear coordinate system are given in the Appendix. Azimuthal symmetry implies $\partial_\theta \approx 0$. One can additionally assume $\partial_\zeta \approx 0$, implying that all physical fields are approximately constant over the GV surface. This is equivalent to a planar approximation applied to a small portion of the GV surface, parameterized by $\tau$ and N, larger than the shock thickness and neglects all the waves propagating in the azimuthal and tangential directions on the GV surface. This can be practically achieved by restricting the analysis to the neighborhood of the origin $\xi \approx 0, \zeta \approx 0$ lying on the GV surface. The conservation equations then become:

$$\partial_\tau \tilde{\rho} + \partial_\xi \left( \tilde{\rho} \tilde{v}_\xi \right) = -\tilde{\rho} \tilde{v}_\xi \mathbb{A} - \tilde{\rho} \tilde{v}_\zeta \mathbb{B} \qquad 20$$

$$\begin{aligned}
&\partial_\tau \left( \tilde{\rho} \tilde{v}_\xi \right) + \partial_\xi \left( \tilde{\rho} \tilde{v}_\xi^2 + \tilde{p} - \tilde{B}_\xi^2 + \tilde{B}_\theta^2 + \tilde{B}_\zeta^2 \right) = -\tilde{\rho} \tilde{v}_\xi \partial_\tau \left( \log \tilde{I}(\tau) \right) \\
&- \mathbb{A} \left( \tilde{\rho} \tilde{v}_\xi^2 - \tilde{B}_\xi^2 + \tilde{B}_\theta^2 + \tilde{B}_\zeta^2 \right) - \mathbb{B} \left( \tilde{\rho} \tilde{v}_\zeta \tilde{v}_\xi - 2 \tilde{B}_\zeta \tilde{B}_\xi \right) - \mathbb{C} \left( \tilde{\rho} \tilde{v}_\xi \tilde{v}_\zeta - 2 \tilde{B}_\xi \tilde{B}_\zeta \right) \\
&+ \mathbb{D} \left( \tilde{\rho} \tilde{v}_\zeta^2 + \tilde{p} + \tilde{B}_\xi^2 + \tilde{B}_\theta^2 - \tilde{B}_\zeta^2 \right) - \mathbb{E} \left( \tilde{\rho} \tilde{v}_\theta^2 + \tilde{p} + \tilde{B}_\xi^2 - \tilde{B}_\theta^2 + \tilde{B}_\zeta^2 \right)
\end{aligned} \qquad 21$$

$$\begin{aligned}
&\partial_\tau \left( \tilde{\rho} \tilde{v}_\theta \right) + \partial_\xi \left( \tilde{\rho} \tilde{v}_\xi \tilde{v}_\theta - 2 \tilde{B}_\xi \tilde{B}_\theta \right) = -\tilde{\rho} \tilde{v}_\theta \partial_\tau \left( \log \tilde{I}(\tau) \right) - \mathbb{A} \left( \tilde{\rho} \tilde{v}_\xi \tilde{v}_\theta - 2 \tilde{B}_\xi \tilde{B}_\theta \right) \\
&- \mathbb{B} \left( \tilde{\rho} \tilde{v}_\zeta \tilde{v}_\theta - 2 \tilde{B}_\zeta \tilde{B}_\theta \right) + \mathbb{E} \left( \tilde{\rho} \tilde{v}_\theta \tilde{v}_\xi - 2 \tilde{B}_\theta \tilde{B}_\xi \right) - \mathbb{F} \left( \tilde{\rho} \tilde{v}_\theta \tilde{v}_\zeta - 2 \tilde{B}_\theta \tilde{B}_\zeta \right)
\end{aligned} \qquad 22$$

$$\begin{aligned}
&\partial_\tau \left( \tilde{\rho} \tilde{v}_\zeta \right) + \partial_\xi \left( \tilde{\rho} \tilde{v}_\xi \tilde{v}_\zeta - 2 \tilde{B}_\xi \tilde{B}_\zeta \right) = -\tilde{\rho} \tilde{v}_\zeta \partial_\tau \left( \log \tilde{I}(\tau) \right) - \mathbb{A} \left( \tilde{\rho} \tilde{v}_\xi \tilde{v}_\zeta - 2 \tilde{B}_\xi \tilde{B}_\zeta \right) \\
&- \mathbb{B} \left( \tilde{\rho} \tilde{v}_\zeta^2 + \tilde{p} + \tilde{B}_\xi^2 + \tilde{B}_\theta^2 - \tilde{B}_\zeta^2 \right) + \mathbb{C} \left( \tilde{\rho} \tilde{v}_\xi^2 + \tilde{p} - \tilde{B}_\xi^2 + \tilde{B}_\theta^2 + \tilde{B}_\zeta^2 \right) \\
&- \mathbb{D} \left( \tilde{\rho} \tilde{v}_\zeta \tilde{v}_\xi - 2 \tilde{B}_\zeta \tilde{B}_\xi \right) + \mathbb{F} \left( \tilde{\rho} \tilde{v}_\theta^2 + \left( \tilde{p} + \tilde{B}_\xi^2 - \tilde{B}_\theta^2 + \tilde{B}_\zeta^2 \right) \right)
\end{aligned} \qquad 23$$

$$\begin{aligned}
&\partial_\tau \tilde{\mathcal{E}} + \partial_\xi \left( \tilde{\mathcal{E}} \tilde{v}_\xi + \tilde{p} \tilde{v}_\xi + 2 \tilde{v}_\xi \tilde{B}^2 - 2 \tilde{B}_\xi \left( \tilde{v} \cdot \tilde{B} \right) \right) \\
&= -\tilde{\mathcal{U}}_L - 2 \tilde{\mathcal{E}} \partial_\tau \log \tilde{I}(\tau) - \left( \tilde{\mathcal{E}} \tilde{v}_\xi + \tilde{p} \tilde{v}_\xi + 2 \tilde{v}_\xi \tilde{B}^2 - 2 \tilde{B}_\xi \left( \tilde{v} \cdot \tilde{B} \right) \right) \mathbb{A} \\
&- \left( \tilde{\mathcal{E}} \tilde{v}_\zeta + \tilde{p} \tilde{v}_\zeta + 2 \tilde{v}_\zeta \tilde{B}^2 - 2 \tilde{B}_\zeta \left( \tilde{v} \cdot \tilde{B} \right) \right) \mathbb{B}
\end{aligned} \qquad 24$$

$$\partial_\tau \tilde{B}_\xi = -\tilde{B}_\xi \partial_\tau \log \tilde{I}(\tau) + (\mathbb{B} - \mathbb{C})\left( \tilde{v}_\xi \tilde{B}_\zeta - \tilde{v}_\zeta \tilde{B}_\xi \right) \qquad 25$$

$$\begin{aligned}
&\partial_\tau \tilde{B}_\theta + \partial_\xi \left( \tilde{v}_\xi \tilde{B}_\theta - \tilde{B}_\xi \tilde{v}_\theta \right) = -\tilde{B}_\theta \partial_\tau \log \tilde{I}(\tau) \\
&- (\mathbb{A} + \mathbb{E})\left( \tilde{v}_\xi \tilde{B}_\theta - \tilde{v}_\theta \tilde{B}_\xi \right) + (\mathbb{F} - \mathbb{B})\left( \tilde{v}_\zeta \tilde{B}_\theta - \tilde{v}_\theta \tilde{B}_\zeta \right)
\end{aligned} \qquad 26$$

$$\partial_\tau \tilde{B}_\zeta + \partial_\xi \left( \tilde{v}_\xi \tilde{B}_\zeta - \tilde{B}_\xi \tilde{v}_\zeta \right) = -\tilde{B}_\zeta \partial_\tau \log \tilde{I}(\tau) + (\mathbb{D} - \mathbb{A})\left( \tilde{v}_\xi \tilde{B}_\zeta - \tilde{v}_\zeta \tilde{B}_\xi \right) \qquad 27$$

$$\tilde{\nabla} \cdot \tilde{B} = \partial_\xi \tilde{B}_\xi + \tilde{B}_\xi \mathbb{A} + \tilde{B}_\zeta \mathbb{B} = 0 \qquad 28$$

Seven out of the 9 equations 20-28 have the form of one dimensional hyperbolic conservation laws with geometric and time-dependent source terms [48]; equations 25 and 28 do not have such forms. All the geometric source terms are proportional to coefficients $\mathbb{A}, \cdots, \mathbb{F}$, (see Appendix) which are related to the curvature of the local coordinate system; the time-dependent source term is related to the scaled current profile determined from the resistive GV model [23]. An indication of their influence can be seen from equations 25 and 27: even a small ambient field like the earth's magnetic field [7,49] can lead to growth of magnetic field components other than the azimuthal magnetic field driving the plasma.

A fractional time-step method [48], constructing a partial solution of homogeneous equations and another partial solution of inhomogeneous equations without advection at alternate half time-steps, is usually utilized for numerical solution of systems of hyperbolic conservation law equations with source terms; this, however, requires development of a Riemann solver [48] for these equations which is outside the scope of this work. This paper is therefore limited to a demonstration that at every fractional time step, the Rankine-Hugoniot conditions [48] for the homogeneous system of hyperbolic conservation laws, yield expressions for the axial magnetic field proportional to the driving azimuthal magnetic field and three components of fluid velocity with kinetic energy density comparable with magnetic energy density in the post-shock state, *which therefore cannot be neglected in theoretical models of DPF* [9-20].

## V. Rankine-Hugoniot conditions and their consequences:

The main thrust of this discussion is to see how transverse flow and magnetic field components other than the azimuthal component are related to the shock propagation. For this purpose, the Rankine-Hugoniot jump conditions [48] across the shock moving with dimensionless speed $C_s$ into the initially stationary ($\tilde{v}_0 = 0$) and almost magnetic-field-free ($\tilde{B}_0 \approx 0$) gas are written below with subscripts 0 and 1 denoting quantities ahead of (in the unperturbed region) and behind shock

$$C_s(\tilde{\rho}_1 - \tilde{\rho}_0) = \tilde{\rho}_1 \tilde{v}_{\xi 1} \qquad 29$$

$$C_s \tilde{\rho}_1 \tilde{v}_{\xi 1} = \left(\tilde{\rho}_1 \tilde{v}_{\xi 1}^2 + \tilde{p}_1 - \tilde{B}_{\xi 1}^2 + \tilde{B}_{\theta 1}^2 + \tilde{B}_{\zeta 1}^2\right) - \tilde{p}_0 \qquad 30$$

$$C_s \tilde{\rho}_1 \tilde{v}_{\theta 1} = \tilde{\rho}_1 \tilde{v}_{\xi 1} \tilde{v}_{\theta 1} - 2\tilde{B}_{\xi 1} \tilde{B}_{\theta 1} \qquad 31$$

$$C_s \tilde{\rho}_1 \tilde{v}_{\zeta 1} = \tilde{\rho}_1 \tilde{v}_{\xi 1} \tilde{v}_{\zeta 1} - 2\tilde{B}_{\xi 1} \tilde{B}_{\zeta 1} \qquad 32$$

$$C_s \left( \tilde{\rho}_1 \tilde{\varepsilon}_{eff} + \frac{\tilde{p}_1}{(\gamma_1 - 1)} + \frac{1}{2}\tilde{\rho}_1 \tilde{v}_1^2 + \tilde{B}_1^2 - \frac{\tilde{p}_0}{(\gamma_0 - 1)} \right)$$
$$= \left( \left( \tilde{\rho}_1 \tilde{\varepsilon}_{eff} + \frac{\tilde{p}_1}{(\gamma_1 - 1)} + \frac{1}{2}\tilde{\rho}_1 \tilde{v}_1^2 + \tilde{B}_1^2 \right) \tilde{v}_{\xi 1} + \tilde{p}_1 \tilde{v}_{\xi 1} + 2\tilde{v}_{\xi 1} \tilde{B}_1^2 - 2\tilde{B}_{\xi 1} (\tilde{v}_1 \cdot \tilde{B}_1) \right)$$ 33

$$C_s \tilde{B}_{\theta 1} = \tilde{v}_{\xi 1} \tilde{B}_{\theta 1} - \tilde{B}_{\xi 1} \tilde{v}_{\theta 1}$$ 34

$$C_s \tilde{B}_{\zeta 1} = \tilde{v}_{\xi 1} \tilde{B}_{\zeta 1} - \tilde{B}_{\xi 1} \tilde{v}_{\zeta 1}$$ 35

Introducing the specific volume $\tilde{V} \equiv \tilde{\rho}^{-1}$, equation 29 becomes

$$\tilde{v}_{\xi 1} = C_s \left( 1 - \tilde{V}_1 \right)$$ 36

Equations 36, 31 and 34 give

$$\tilde{B}_{\xi 1}^2 = \frac{1}{2} \tilde{V}_1 C_s^2$$ 37

$$\tilde{v}_{\theta 1} = -\sqrt{2\tilde{V}_1} \tilde{B}_{\theta 1}$$ 38

Equations 36, 32 and 35 give

$$\tilde{v}_{\zeta 1} = -\sqrt{2\tilde{V}_1} \tilde{B}_{\zeta 1}$$ 39

From 28 and 37, tangential component of post-shock magnetic field is obtained as,

$$\tilde{B}_{\zeta 1} = -\tilde{B}_{\xi 1} \mathbb{Q} \qquad \text{with } \mathbb{Q} \equiv \mathbb{A}/\mathbb{B} \text{ given in the Appendix}$$ 40

The Rayleigh Line (RL) is derived from 29 and 30

$$\frac{\tilde{p}_1 - \tilde{p}_0 + \left( \tilde{B}_{\theta 1}^2 + \tilde{B}_{\zeta 1}^2 - \tilde{B}_{\xi 1}^2 \right)}{\left( \tilde{V}_1 - 1 \right)} = -C_s^2$$ 41

and, using 37 and 40, is further transformed as

$$\frac{\tilde{p}_1 - \tilde{p}_0 + \tilde{B}_{\theta 1}^2}{\frac{1}{2}\mathbb{Y}\tilde{V}_1 - 1} = -C_s^2 \qquad \text{where } \mathbb{Y} \equiv (1+\mathbb{Q}^2) = \tilde{r}^2/N^2 \qquad\qquad 42$$

For the shock to be locally unidirectional on the entire GV surface, the shock velocity must be proportional to the normal sheath velocity given by equation 1 *everywhere on the GV surface*: $C_s = C\tilde{B}_{\theta 1}$, where C, the velocity parameter, is a proportionality constant that is independent of N or $\tilde{r}$. This simply means that the shock wave corresponding to the "effective magnetic piston" represented by a GV surface at dimensionless time $\tau$ is located at another GV surface associated with $\tau' = C\tau$. Using the driving azimuthal magnetic field in scaled form $\tilde{B}_{\theta 1} = \tilde{r}^{-1}$, the Rayleigh Line becomes

$$\tilde{p}_1 = \tilde{p}_0 + (C^2 - 1)\tilde{r}^{-2} - \tfrac{1}{2}C^2\tilde{V}_1/N^2 \qquad\qquad 43$$

For $\tilde{p}_1 > \tilde{p}_0$, 43 shows that C must be greater than unity. The Hugoniot is obtained from 33 and previous results as

$$\left(\frac{\gamma_1\tilde{p}_1\tilde{V}_1}{(\gamma_1-1)} - \frac{\tilde{p}_0\gamma_0}{(\gamma_0-1)}\right) - (\tilde{p}_1-\tilde{p}_0)\left\{\frac{\tilde{V}_1\mathbb{Y}(\tfrac{3}{2}-\tilde{V}_1) - \tfrac{3}{2}}{\tfrac{1}{2}\mathbb{Y}\tilde{V}_1 - 1}\right\} - \tilde{B}_{\theta 1}^2\frac{(\tilde{V}_1-1)\tilde{V}_1\mathbb{Y} + \tfrac{1}{2}}{\tfrac{1}{2}\mathbb{Y}\tilde{V}_1 - 1} + \tilde{\varepsilon}_{\text{eff}} = 0 \qquad 44$$

This Hugoniot, like the one for the case of detonation, does not pass through the initial state $(\tilde{p}_0, 1)$ in the $(\tilde{p}, \tilde{V})$ space. *Unlike the case of detonation, even the Rayleigh Line does not pass through the initial state*.

Note that the sign of $\tilde{B}_{\xi 1}$ and hence of $\tilde{B}_{\zeta 1}$, remains undetermined from 37. The post-shock magnetic field is given by:

$$\begin{aligned}
\vec{B}/B_0 &= -\tilde{B}_{\xi 1}\mathbb{Q}\hat{\zeta} + \tilde{B}_{\xi 1}\hat{\xi} + \tilde{B}_{\theta_1}\hat{\theta} \\
&= -\tilde{B}_{\xi_1}\mathbb{Q}\left(s\tilde{r}^{-1}\sqrt{\tilde{r}^2-N^2}\hat{z} + \tilde{r}^{-1}N\hat{r}\right) + \tilde{B}_{\xi_1}\left(-s\tilde{r}^{-1}\sqrt{\tilde{r}^2-N^2}\hat{r} + \tilde{r}^{-1}N\hat{z}\right) + \tilde{r}^{-1}\hat{\theta} \qquad 45 \\
&= \pm\hat{z}CN^{-1}\sqrt{\tilde{V}_1/2} + \hat{\theta}\tilde{r}^{-1}
\end{aligned}$$

This relation shows that existence of an axial magnetic field of undetermined polarity, proportional to $B_0$, related to the curved geometry of the sheath (through the factor $N^{-1}$) and to the post-shock density [30], is predicted by the conservation laws. The ratio of the axial magnetic field to azimuthal magnetic field in the radial implosion phase increases near the anode because of the factor $N^{-1}$. *Within the ambit of the approximations used above*, it does not reverse its sign within the region occupied by the plasma; relation 45 therefore predicts emission of magnetic flux from the DPF [31]. The question of polarity of the axial magnetic field has been addressed elsewhere [7,25].

The post-shock velocity vector is given by

$$\vec{v}_1/v_A = \tilde{B}_{\theta 1}\left(-\pm sN^{-1}\sqrt{\tilde{r}^2 - N^2}\tilde{V}_1 C\hat{\zeta} + C\left(1 - \tilde{V}_1\right)\hat{\xi} - \sqrt{2\tilde{V}_1}\hat{\theta}\right) \qquad 46$$

Note that the tangential component of velocity in 46 changes sign at the N=1 turn-around characteristic in fig 5: the region to its right side is a continuation of the rundown region for which s=-1 [23], while on its left, s=+1. The "toroidal directionality" indicated by the data cited in Section II refers to the pinch phase, which is outside the limit of validity of the GV model and hence of this discussion; its origin is, however, seen to be contained in the conservation laws formulated above.

The post-shock azimuthal velocity 38 is seen to be associated with ion kinetic energy equal to magnetic energy per particle: $\tfrac{1}{2}m_i v_{\theta 1}^2 = B_{\theta 1}^2/2\mu_0 n_1$. For post-shock magnetic field ~ 200 T, and density ~$10^{18}$ cm$^{-3}$, this of the order of 100 keV. The ion kinetic energy associated with other than azimuthal directions is also seen to be of the order of the magnetic energy per particle

$$\tfrac{1}{2}m_i v_A^2\left(\tilde{v}_{\xi 1}^2 + \tilde{v}_{\zeta 1}^2\right) = \frac{1}{2}\frac{B_0^2 \tilde{B}_{\theta 1}^2}{2\mu_0 n_0}C^2\left(1 - 2\tilde{V}_1 + \tilde{V}_1^2 \tilde{r}^2/N^2\right) \qquad 47$$

Because their azimuthal motion keeps them confined within the plasma, these energetic ions, mixed with shock wave reflected from the axis, could be expected to lead to the much-higher-than-thermal fusion reaction rate [3]. The observation [28] of ~100 keV deuterons circulating around the axis in Frascati DPF thus has a natural explanation in terms of conservation laws without invoking any m=0 instability, anomalous resistivity or reconfiguration of the magnetized plasma. Indeed, there is evidence from three large facilities about neutron emission being unaffected by absence of m=0 instability [50-52].

The post-shock electric field, which is also equal to the electric field ahead of shock because of electromagnetic boundary conditions [45], is seen to contain an azimuthal component:

$$\vec{E}_1 = -\vec{v}_1 \times \vec{B}_1$$

$$= -\frac{B_0^2}{\sqrt{2\mu_0 \rho_0}} \left( +\hat{\zeta}(\tilde{v}_{\theta 1} B_{\xi 1} - \tilde{v}_{\xi 1} B_{\theta 1}) + \hat{\theta}(\tilde{v}_{\xi 1} B_{\zeta 1} - \tilde{v}_{\zeta 1} B_{\xi 1}) + \hat{\xi}(\tilde{v}_{\zeta 1} B_{\theta 1} - \tilde{v}_{\theta 1} B_{\zeta 1}) \right) \qquad 48$$

$$= \frac{B_0^2}{\sqrt{2\mu_0 \rho_0}} \tilde{B}_{\theta 1}^2 \left( \hat{\zeta} C \pm \hat{\theta} \mathbb{Q} C^2 \sqrt{\tilde{V}_1/2} \right)$$

In the region outside the squirrel-cage cathode, the azimuthal electric field is governed by Maxwell's equations in vacuum, with the boundary conditions that it should equal the azimuthal electric field in 48 at $\tilde{r}_C$ and should be zero at the metallic vacuum chamber. Surface currents at the metallic vacuum chamber, therefore, should cause reversal of the axial magnetic flux in the region between the cathode and the vacuum chamber.

An interesting aspect of these results is that they suggest a helical structure for the Poynting vector. This may possibly explain an unusual observation [54] in a small plasma focus. Cone-shaped copper foil placed on the anode was found to be *twisted*, rather than crumpled, after the shot, which is possible only if the electromagnetic force acting on the foil had a helical structure.

The above exercise does not represent a complete solution to the problem of the structure and propagation of the plasma current sheath in the DPF. Additional physics, possibly related to ionization dynamics in the region ahead of shock[45], needs to be formulated to determine the velocity parameter C in a consistent manner. Implementation of a Riemann solver with fractional time step method is necessary for taking into account the geometric and time-dependent source terms, which would probably make C a function of $\tau$. It is also not clear at present whether such complete solution can be obtained within the limits of the GV model; the assumption that the shock propagation remains unidirectional in the curvilinear coordinate system constructed using the GV model may not remain valid while incorporating ionization dynamics and fractional time step method so that either tangential variation of physical quantities may have to be taken into account in a two-dimensional hyperbolic conservation law system or the coordinate system itself may have to evolve with dimensionless time maintaining one dimensional character of shock propagation.

*In spite of these limitations, the above discussion achieves the aim of demonstrating that axial magnetic field proportional to the driving azimuthal magnetic field and 3-D plasma flow with toroidal directionality and high ion kinetic energy are natural consequences of conservation laws in the curved axisymmetric geometry of the PCS, and therefore, they should not be neglected in theoretical models of the DPF.*

### VI. Summary and conclusions:

This paper demonstrates that axial magnetic field and toroidally streaming fast ions in the dense plasma focus are natural consequences of conservation laws in the curved axisymmetric geometry of the current sheath. This demonstration uses a curvilinear coordinate system constructed using the Gratton-Vargas model, which captures the curved shape of the sheath using analytical formula, in which, the shock propagation can be treated as locally unidirectional. Equations for conservation of mass, momentum and energy, including contribution from electromagnetic fields and advection of magnetic field lead to a system of seven one-dimensional hyperbolic conservation law equations with geometric and time-dependent source terms under the assumption that physical fields have significant variation only along the normal to the shock wave. Expressions for three components of fluid velocity, axial component of magnetic field and azimuthal component of electric field arise naturally in the system of Rankine-Hugoniot jump conditions.

The significance of this exercise is that no anomalous phenomena or instabilities need to be invoked to explain the presence ~100 keV ions confined within the plasma: the entire post-shock state acquires kinetic energy of this order as it approaches the pinch state magnetic intensity ~ 200 T at a density of $10^{18}$ cm$^{-3}$. This, however, is the average, not thermal, velocity of the fluid. Reflection of the shock from the axis can be expected to cause a high-relative-velocity distribution for ion pairs, resulting in a much-larger-than-thermal velocity-averaged reaction rate, explaining the main feature of interest about the dense plasma focus. Shifted-Maxwellian velocity distributions exhibit inverse Landau damping, consistent with observations of turbulence [28,39] in the post-shock state.

This theory is supported by experimental evidence regarding existence of axial magnetic field and fast ions having toroidal directionality cited in Section II; it also enables design of further confirmatory diagnostic experiments. One such diagnostic is the measurement of azimuthal electric field between the cathode and the vacuum chamber, which is a unique prediction of this theory, using a series of concentric planar diamagnetic loops. By fitting the measured radial and temporal variation to solutions of Maxwell's equations in vacuum, with metallic boundary at the vacuum wall, one can obtain the value of the azimuthal electric field at the cathode radius, which can be compared with Equation 48, with C determined from the tangential potential drop measured between anode and cathode using a coaxial voltage probe [55] and $\tilde{V}_1$ calculated from Equations 43 and 44.

This exercise may have wider implications: shock propagation in other curved axisymmetric geometries (such as axially non-uniform cylindrical geometry for a gas-puff z-pinch or spherical geometry as in inertial confinement fusion or star formation), with physical fields predominantly varying along one direction, may exhibit similar results. It could even be a candidate mechanism for origin of magnetic field in stars.

Its major implication, however, should be re-examination of the possibility of the dense plasma focus, designed to work at high pressure, serving as an inexpensive means of generating nuclear reactions for medical isotope production or production of fusion energy from p-$^{11}$B reactions, using sub-MeV fast ions continuously generated and trapped within the plasma without any additional hardware.

## VII. Acknowledgements:

References

1. J. W. Mather, "Dense Plasma Focus", Methods of Experimental Physics, vol. 9B, p. 187, 1971
2. M. Krishnan, "The Dense Plasma Focus: A Versatile Dense Pinch for Diverse Applications" IEEE TRANS. PLASMA SCI., VOL. 40, (2012) p. 3189.
3. G. Bockle, J. Ehrhardt, P. Kirchesch, N. Wenzel., R. Batzner, H. Hinsch and K. Hübner, "Spatially resolved light scattering diagnostic on plasma focus devices", Plasma Phys. Cont Fusion, 34, pp 801-841, 1992
4. J.S. Brzosko and V. Nardi, "High yield of $^{12}C(d,n)^{13}N$ and $^{14}N(d,n)^{15}O$ reactions in the plasma focus pinch", Physics Letters A, 155, (1991) p.162.
5. J.S. Brzosko, V. Nardi, J.R. Brzosko, D. Goldstein, "Observation of plasma domains with fast ions and enhanced fusion in plasma-focus discharges" Physics Letters A 192 250-257 (1994).
6. Jan S. Brzosko, Krzysztof Melzacki, Charles Powell, Moshe Gai, Ralph H., France III , James E. McDonald , Gerald D. Alton, Fred E. Bertrand, and James R. Beene, "Breeding 1010/s Radioactive Nuclei in a Compact Plasma Focus Device", AIP Conf. Proc. **576**, 277 (2001); doi: 10.1063/1.1395303.
7. S K H Auluck, "Dense Plasma Focus: A question in search of answers, a technology in search of applications" Plasma Science and Applications (ICPSA 2013) International Journal of Modern Physics: Conference Series, Vol. 32 (2014) 1460315.
8. A. Bernard et. al. "Scientific status of the dense plasma focus", J Moscow Phys. Soc., 8, (1998), 93-170
9. S. Lee, "Plasma Focus Radiative Model: Review of the Lee Model Code," J Fusion Energy, August 2014, Volume 33, Issue 4, pp 319-335, doi: 10.1007/s10894-014-9683-8.
10. D. E. Potter. "Numerical studies of the plasma focus" Phys.Fluids 14(9) (1971), pp. 1911–1925.
11. S. Maxon and J. Eddleman, "Two dimensional magnetohydrodynamic calculations of the plasma focus" Phys. Fluids, 21, 1856 (1978).
12. S. F. Garanin and V. I. Mamyshev, "Two-Dimensional MHD Simulations of a Plasma Focus with Allowance for the Acceleration Mechanism for Neutron Generation" Plasma Physics Reports, Vol. 34, No. 8, pp. 639–649. (2008).
13. P. G. Elgorth, "Comparison of plasma focus calculations" Phys. Fluids 25, 2408 (1982);
14. B. T. Meehan, J. H. J. Niederhaus, "Fully Three-dimensional Simulation and Modeling of a Dense Plasma Focus" arXiv:1402.5083v1 [physics.plasm-ph] 20 Feb 2014
15. K. V. Roberts and D. E. Potter, Methods of Computational Physics. New York: Academic, (1970).
16. Casanova F, Correa G, Moreno C and Clausse A, "Experimental study and modeling of the plasma dynamics of magnetically driven shock waves in a coaxial tube," Plasma Phys. Control. Fusion 45 (2003) 1989–99;
Casanova F, Moreno C and Clausse A, "Finite-elements numerical model of the current sheet movement and shaping in coaxial discharges," Plasma Phys. Control. Fusion 47 (2005) 1239–50.
17. Stepniewski W, "MHD numerical modeling of the plasma focus phenomena," Vacuum (2004) 7651–55 .

18. González J, Florido P, Bruzzone H and Clausse A "A lumped parameter model of plasma focus," IEEE Trans. Plasma Sci 32 (2004) 1383–91.
19. González J H, Brollo F R and Clausse A, " Modeling of the dynamic plasma pinch in plasma focus discharges based in Von Karman approximations", IEEE Trans. Plasma Sci 37 (2009) 2178-85.
20. A. Schmidt, V. Tang, D. Welch, "Fully Kinetic Simulations of Dense Plasma Focus Z-Pinch Devices," Phys. Rev. Letters, vol. 109, 205003 (2012).
21. V. I. Krauz, K. N. Mitrofanov, M. Scholz, M. Paduch, P. Kubes, L. Karpinski, and E. Zielinska "Experimental evidence of existence of the axial magnetic field in a plasma focus", Eur Phy. Lett, vol 98 (2012) 45001

22. P Kubes, V Krauz, K Mitrofanov, M Paduch, M Scholz, T Piszarzcyk, T. Chodukowski, Z Kalinowska, L Karpinski, D Klir, J Kortanek, E Zielinska, J Kravarik and K Rezac, "Correlation of magnetic probe and neutron signals with interferometry figures on the plasma focus discharge" Plasma Phys. Control. Fusion 54 (2012) 105023.

23. S K H. Auluck, "Re-appraisal and extension of the Gratton-Vargas two-dimensional analytical snowplow model of plasma focus evolution in the context of contemporary research" PHYSICS OF PLASMAS 20, 112501 (2013).
24. F. Gratton and J.M. Vargas, "Two dimensional electromechanical model of the plasma focus", in Energy Storage, Compression and Switching, V. Nardi, H. Sahlin, and W. H. Bostick, Eds., vol. 2. New York: Plenum, 1983, p. 353.

25. V Krauz, K Mitrofanov, M Scholz, M Paduch, L Karpinski, E Zielinska and P Kubes, "Experimental Study of the Structure of the Plasma-Current Sheath on the PF-1000 Facility", Plasma Phy. Cont. Fus. 54 (2012) , 025010

26. S.K.H. Auluck, "Role of Electron-inertia-linked Current Source Terms in the Physics of Cylindrically Symmetric Imploding Snow-Plow Shocks", Phys. of Plasmas, 9, (2002) 4488-94

27. V.A. Gribkov, A.V. Dubrovsky, N.V. Kalachev, T. A. Kozlova and V.Ya. Nikulin, "Dynamics of plasma phenomena in "plasma focus" under the action of powerful laser radiation", http://jphyscol.journaldephysique.org/articles/jphyscol/pdf /1979/07/physcol197940 C7369.pdf.
28. V.A. Gribkov, A.V. Dubrovsky, N.V. Kalachev, T. A. Kozlova and V.Ya. Nikulin. "Dynamics of plasma phenomena in "plasma focus" under the action of powerful laser radiation", Bulletin of Lebedev Physical Institute (1980) vol. 127, pp. 32 - 61 (in Russian)
29.
30. J.P. Rager, "Progresses on Plasma Focus research at Frascati", Invited paper at the 10th European Conference on Plasma Physics and CTR", Frascati Report 81.43/cc.
31. U. Jäger and H. Herold, "Fast ion kinetics and fusion reaction mechanism in the plasma focus" Nucl. Fus. 27 407 (1987).
32. S.K.H. AULUCK, "An apparent momentum balance anomaly in the Plasma Focus." IEEE Trans. Plasma. Sci, vol.25, p.37, 1997.

33. S.K.H. AULUCK, "Manifestation of Constrained Dynamics in a Low-Pressure Spark", IEEE Trans. Plasma Sci., Vol. 41, No. 3, March 2013 pp. 437-446.
34. M. J. Bernstein, D. A. Mescan, and H. L. L. Van Passen, "Space, time and energy distribution of Neutrons and X-rays from a focussed plasma discharge," *Phys. Fluids* vol. 12, pp. 2193–2202, 1969.
35. J. H. Lee, L. P. Shomo, M. D. Williams, and H. Hermansdorfer, "Neutron production mechanism in a plasma focus," *Phys. Fluids*, vol. 14, pp. 2217–2223, 1971.
36. P. Kubes, J. Kravarik, D. Klir, K. Rezac, M. Scholz, M. Paduch, K. Tomaszewski, I. Ivanova-Stanik, B. Bienkowska, L. Karpinski, M. J. Sadowski, and H. Schmidt, "Correlation of Radiation With Electron and Neutron Signals Taken in a Plasma-Focus Device" *IEEE Trans. Plasma Science*, Vol. 34, pp. 2349-2355, 2006.
37. Ch. Maisonnier, C. Gourlan, G. Luzzi, L. Papagno, F. Peccorella, J. P. Rager, B. V. Robouch, and P. Samuelli, "Structure of the dense plasma focus–Part II: Neutron measurements and phenomenological description," *Plasma Phys. Cont. Nucl. Fusion*, vol. 1, pp. 523–535, 1971.
38. M. J. Bernstein and G. G. Comisar, "Neutron energy and flux distributions from a crossed-field acceleration model of plasma focus and Z-pinch discharges", *Phys. Fluids* 15, p.700-, 1972.
39. H. Schmidt, A. Kasperczuk, M. Paduch, T. Pisarczyk, M. Scholz, K.Tomaszewski and A. Szydłowski,"Review of Recent Experiments with the Megajoule PF-1000 Plasma Focus Device", *Physica Scripta*. Vol. 66, 168-172, 2002
40. M. V. Roshan, P. Lee, S. Lee, A. Talebitaher, R. S. Rawat, and S. V. Springham, "Backward high energy ion beams from plasma focus" *Physics Of Plasmas* vol. 16, pp 074506- (2009).
41. A. Bernard, A. Coudeville, A. Jolas, J. Launspach, and J. Mascureau, "Experimental studies of the plasma focus and evidence for nonthermal processes," *Phys. Fluids*, vol. 18, pp. 180–194, 1975.
42. M.M. Milanese, J.O. Pouzo, "Evidence of non-thermal processes in a 1-MJ plasma focus device by analyzing the neutron spectra", *Nucl. Fusion*, vol. 18, pp.533-536, 1978.
43. U. Jaeger, Ph.D. Thesis, Institut Fuer Plasmaforschung, University of Stuttgart, Report-IPF-86-1, 1986.

44. S. K. H. Auluck, "Evaluation of Turner relaxed state as a model of long-lived ion-trapping structures in plasma focus and Z-pinches" *Physics of. Plasmas*, vol 18, 032508, 2011.
45. P Kubes, D Klir, J Kravarik, K Rezac, J Kortanek, V Krauz, K Mitrofanov, M Paduch, M Scholz, T Pisarczyk, T Chodukowski, Z Kalinowska, L Karpinski and E Zielinska, "Scenario of pinch evolution in a plasma focus discharge" , Plasma Phys. Control. Fusion **55** (2013) 035011 (8pp).
46. S K H Auluck, "Bounds imposed on the sheath velocity of a dense plasma focus by conservation laws and ionization stability condition" Physics of Plasmas, **21**, 090703 (2014); doi: 10.1063/1.4894680
47. C.R. Haas, R. Noll, F. Rohl, G. Herziger, "Schlieren diagnostics of the plasma focus" Nucl. Fusion 24 (1984) 1216-1220.
48. J. M. Vargas, F. Gratton, J. Gratton, H. Bruzzone, and H. Kelly, "Experimental verification of a theory of the current sheath in the plasma focus." in Proc. 6th Int. Conf. on Plasma Phys. and Controlled Fusion Res., Berchtesgaden, 1976, IAEA-CN-35/E18.5(a), p. 483.
49. S K H Auluck "New insights from the resistive Gratton-Vargas model", Presented at the International Workshop and Expert Meeting of the ICDMP, October 2014.

50. Randall J. LeVeque, "Nonlinear Conservation Laws And Finite Volume Methods For Astrophysical Fluid Flow" in Computational Methods for Astrophysical Fluid Flow, Saas-Fee Advanced Course 27. Lecture Notes 1997, Swiss Society for Astrophysics and Astronomy, LeVeque, R.J., Mihalas, D., Dorfi, E.A., Müller, E. Steiner, Oskar, Gautschy, A. (Eds.) ISBN 978-3-540-31632-9; Randall J. LeVeque, "Finite Volume Methods for Hyperbolic Problems", Cambridge Univesity Press, 2004, ISBN 0-511-04219-1

51. J . W. Mather and H. S. Ahluwalia, "The Geomagnetic Field-An Explanation for the Microturbulence in Coaxial Gun Plasmas", IEEE TRANS. PLASMA SCIENCE, 16. (1988) p. 56.

52. R. Schmidt, Ph.D. dissertation, Univ. Stuttgart., 1986.

53. J. P. Rager, "The plasma focus," in *Unconventional Approaches to Fusion*, B. Brunnelli and G. G. Leotta, Eds. New York: Plenum, 1981, pp. 157–207.

54. V. A. Bahilov, M. G. Belkov, P. A. Belyaev, I. V. Volubev, V. A. Gribkov, A. V. Dubrovsky, V. M. Zaytsev, Yu. F. Igonin, A. I. Isakov, N. V. Kalachev, E. D. Korop, O. N. Krokhin, I. S. G. Kuznetsov, Yu. V. Marakov, and V. Ya. Nikulin, "Experimental investigations on 'PLAMYA' installation," presented at the 4th Int. Workshop on Plasma Focus and Z-Pinch Res., Warsaw, Sept. 9–11, 1985.

55. V. Ya. Nikulin (private communication) reported in S K H Auluck, "Interferometric diagnostic of plasma rotational flux" "ICDMP Workshop and Expert Meeting On Dense Magnetized Plasmas" Warsaw, Poland, 3-6 December 2007.
http://www.icdmp.pl/phocadownload/2007/auluck.pdf

56. J. W. Mather and P. J. Bottoms, "Characteristics of the Dense Plasma Focus Discharge" Phys. Fluids, 11, 611 (1968); doi: 10.1063/1.1691959

Appendix

Some properties of the $(\zeta, \theta, \xi)$ coordinate system

Equate the line element in the two coordinate systems

$$\hat{r}d\tilde{r} + \hat{z}d\tilde{z} = \hat{\xi}d\xi + \hat{\zeta}d\zeta$$

$$= \hat{r}\left(-d\xi s\tilde{r}^{-1}\sqrt{\tilde{r}^2 - N^2} + d\zeta N\tilde{r}^{-1}\right) + \hat{z}\left(d\xi N\tilde{r}^{-1} + d\zeta s\tilde{r}^{-1}\sqrt{\tilde{r}^2 - N^2}\right)$$

Equating r and z components

$$d\tilde{r} = -d\xi s\tilde{r}^{-1}\sqrt{\tilde{r}^2 - N^2} + d\zeta N\tilde{r}^{-1}; d\tilde{z} = d\xi N\tilde{r}^{-1} + d\zeta s\tilde{r}^{-1}\sqrt{\tilde{r}^2 - N^2}$$

Solve for $d\xi$, $d\zeta$:

$$d\xi = -d\tilde{r}s\tilde{r}^{-1}\sqrt{\tilde{r}^2 - N^2} + d\tilde{z}N\tilde{r}^{-1}; \quad d\zeta = N\tilde{r}^{-1}d\tilde{r} + s\tilde{r}^{-1}\sqrt{\tilde{r}^2 - N^2}d\tilde{z}$$

In the tangential direction $d\xi = 0$ giving $d\zeta = N^{-1}\tilde{r}d\tilde{r}$.

In the normal direction: $d\zeta = 0$ giving $d\xi = -s\tilde{r}d\tilde{r}/\sqrt{\tilde{r}^2 - N^2}$

The differentiation of unit vectors is given below, using the following expressions:

$$\tilde{r}^2 = 2(\zeta - \zeta_0)^2 + 2S(\zeta - \zeta_0)\sqrt{(\zeta - \zeta_0)^2 - (\xi - \xi_0)^2}; S = -s\text{Sign}[\xi - \xi_0]$$

$$N(\xi, \zeta) = (\zeta - \zeta_0) + S\sqrt{(\zeta - \zeta_0)^2 - (\xi - \xi_0)^2}$$

$$\mathbb{A}(\xi, \zeta) \equiv \tilde{r}^{-1}(\xi, \zeta)\partial_\xi(\tilde{r}(\xi, \zeta)); \mathbb{A} \equiv \mathbb{A}(0,0)$$

$$\mathbb{B}(\xi, \zeta) \equiv \tilde{r}^{-1}(\xi, \zeta)\partial_\zeta(\tilde{r}(\xi, \zeta)); \mathbb{B} \equiv \mathbb{B}(0,0)$$

$$\mathbb{Q} \equiv \mathbb{A}/\mathbb{B} = \frac{d\zeta/d\tilde{r}}{d\xi/d\tilde{r}} = -s\frac{\sqrt{\tilde{r}^2 - N^2}}{N}$$

$$\mathbb{Y} = 1 + \mathbb{Q}^2 = \tilde{r}^2/N^2$$

$$\mathbb{C}(\xi, \zeta) \equiv \frac{s\text{Sign}(\xi - \xi_0)}{2\sqrt{(\zeta - \zeta_0)^2 - (\xi - \xi_0)^2}}; \mathbb{C} \equiv \mathbb{C}(0,0);$$

$$\mathbb{D}(\xi, \zeta) = \frac{s|\xi - \xi_0|}{2(\zeta - \zeta_0)\sqrt{(\zeta - \zeta_0)^2 - (\xi - \xi_0)^2}}; \mathbb{D} \equiv \mathbb{D}(0,0)$$

$$\mathbb{E}(\xi,\zeta) \equiv s\tilde{r}^{-2}(\xi,\zeta)\sqrt{\tilde{r}^2 - N^2}; \mathbb{E} \equiv \mathbb{E}(0,0)$$

$$\mathbb{F}(\xi,\zeta) \equiv N\tilde{r}^{-2}(\xi,\zeta); \mathbb{F} \equiv \mathbb{F}(0,0)$$

$$\partial_\xi \hat{\xi} = -\mathbb{C}(\xi,\zeta)\hat{\zeta}; \quad \partial_\xi \hat{\theta} = 0; \quad \partial_\xi \hat{\zeta} = \mathbb{C}(\xi,\zeta)\hat{\xi};$$

$$\partial_\theta \hat{\xi} = -\tilde{r}(\xi,\zeta)\mathbb{E}(\xi,\zeta)\hat{\theta}; \quad \partial_\theta \hat{\theta} = \tilde{r}(\xi,\zeta)\left(\mathbb{E}(\xi,\zeta)\hat{\xi} - \mathbb{F}(\xi,\zeta)\hat{\zeta}\right); \quad \partial_\theta \hat{\zeta} = \tilde{r}(\xi,\zeta)\mathbb{F}(\xi,\zeta)\hat{\theta};$$

$$\partial_\zeta \hat{\xi} = \hat{\zeta}\mathbb{D}(\xi,\zeta); \quad \partial_\zeta \hat{\theta} = 0; \quad \partial_\zeta \hat{\zeta} = -\mathbb{D}(\xi,\zeta)\hat{\xi}$$

The metric for the local coordinate system can be written, for square of line element $d\sigma^2$, as

$$d\sigma^2 = d\xi^2 + \tilde{r}^2(\xi,\zeta)d\theta^2 + d\zeta^2$$

The scale factors are then $h_1 = 1, h_2 = \tilde{r}(\xi,\zeta), h_3 = 1$.

Differential operators can then be written as

$$\tilde{\nabla}F = \hat{\xi}\partial_\xi F + \hat{\theta}\tilde{r}^{-1}(\xi,\zeta)\partial_\theta F + \hat{\zeta}\partial_\zeta F$$

$$\tilde{\nabla}\cdot\tilde{A} = \tilde{r}^{-1}(\xi,\zeta)\left\{\partial_\xi\left(\tilde{r}(\xi,\zeta)\tilde{A}_\xi\right) + \partial_\theta \tilde{A}_\theta + \partial_\zeta\left(\tilde{r}(\xi,\zeta)\tilde{A}_\zeta\right)\right\}$$
$$= \left\{\partial_\xi A_\xi + \tilde{r}^{-1}(\xi,\zeta)\partial_\theta A_\theta + \partial_\zeta A_\zeta + A_\xi \mathbb{A}(\xi,\zeta) + A_\zeta \mathbb{B}(\xi,\zeta)\right\}$$

$$\tilde{\nabla}\cdot\left(\vec{A}\vec{B}\right) =$$

$$= +\hat{\xi}\begin{pmatrix} +\partial_\xi\left(A_\xi B_\xi\right) + \tilde{r}^{-1}(\xi,\zeta)\partial_\theta\left(A_\theta B_\xi\right) + \partial_\zeta\left(A_\zeta B_\xi\right) \\ +\mathbb{A}(\xi,\zeta)A_\xi B_\xi + \mathbb{B}(\xi,\zeta)A_\zeta B_\xi + \mathbb{C}(\xi,\zeta)A_\xi B_\zeta - \mathbb{D}(\xi,\zeta)A_\zeta B_\zeta + A_\theta B_\theta \mathbb{E}(\xi,\zeta) \end{pmatrix}$$

$$+\hat{\theta}\begin{pmatrix} +\partial_\xi\left(A_\xi B_\theta\right) + \tilde{r}^{-1}(\xi,\zeta)\partial_\theta\left(A_\theta B_\theta\right) + \partial_\zeta\left(A_\zeta B_\theta\right) \\ +\mathbb{A}(\xi,\zeta)A_\xi B_\theta + \mathbb{B}(\xi,\zeta)A_\zeta B_\theta - A_\theta B_\xi \mathbb{E}(\xi,\zeta) + A_\theta B_\zeta \mathbb{F}(\xi,\zeta) \end{pmatrix}$$

$$+\hat{\zeta}\begin{pmatrix} +\partial_\xi\left(A_\xi B_\zeta\right) + \tilde{r}^{-1}(\xi,\zeta)\partial_\theta\left(A_\theta B_\zeta\right) + \partial_\zeta\left(A_\zeta B_\zeta\right) \\ +\mathbb{A}(\xi,\zeta)A_\xi B_\zeta + \mathbb{B}(\xi,\zeta)A_\zeta B_\zeta - \mathbb{C}(\xi,\zeta)A_\xi B_\xi + \mathbb{D}(\xi,\zeta)A_\zeta B_\xi - A_\theta B_\theta \mathbb{F}(\xi,\zeta) \end{pmatrix}$$